\documentclass[letterpaper, 10 pt, conference]{ieeeconf}  % Comment this line out if you need a4paper

\IEEEoverridecommandlockouts                              % This command is only needed if 
\usepackage{color}
\usepackage[colorinlistoftodos]{todonotes}
\usepackage[colorlinks=true, allcolors=blue]{hyperref}
\usepackage{ifthen}
\newboolean{showcomments}
\setboolean{showcomments}{true}
\newcommand{\antonis}[1]{\ifthenelse{\boolean{showcomments}}
{ \textcolor{red}{(Antonis says:  #1)}}{}}

\usepackage[style=ieee]{biblatex}
\usepackage{color}

\usepackage{graphics} % for pdf, bitmapped graphics files
\usepackage{epsfig} % for postscript graphics files
\usepackage{amsmath} % assumes amsmath package installed
\usepackage{amssymb}  % assumes amsmath package installed
\usepackage{graphicx}
\usepackage{breqn}
\usepackage{german}
\usepackage{caption}
\usepackage{subcaption}

\usepackage{algorithm}
\usepackage{algpseudocode}

\newtheorem{theorem}{Theorem}
\newtheorem{definition}{Definition}
\newtheorem{proposition}{Proposition}

\title{\LARGE \bf
Stability of Non-linear Neural Feedback Loops using Sum of Squares
}

\author{Matthew Newton and Antonis Papachristodoulou% <-this % stops a space
\thanks{This work was supported by EPSRC grants EP/L015897/1 (to M. Newton) and EP/M002454/1 (to A. Papachristodoulou).}% <-this % stops a space
\thanks{M. Newton and A. Papachristodoulou are with the Department of Engineering Science, University of Oxford, Parks Road, Oxford, OX1 3PJ, U.K. {\tt\small \{matthew.newton,antonis\}@eng.ox.ac.uk}}%
}

\begin{document}

\maketitle
\thispagestyle{empty}
\pagestyle{empty}

\begin{abstract}
Neural network controllers have the potential to improve the performance of feedback systems compared to traditional controllers, due to their ability to act as general function approximators. However, quantifying their safety and robustness properties has proven challenging due to the non-linearities of the activation functions inside the neural network. A key robustness indicator is certifying the stability properties of the feedback system and providing a region of attraction, which has been addressed in previous literature. However, these works only address linear systems or require one to abstract the plant non-linearities and bound them using slope and sector constraints. In this paper we use a Sum of Squares programming framework to compute the stability of non-linear systems with neural network controllers directly. Within this framework, we can propose higher order candidate Lyapunov functions with richer structures that are able to better capture the dynamics of the non-linear system and the nonlinearities in the neural network. We are also able to analyse these systems in continuous time, whereas other methods rely on discretising the system. These higher order Lyapunov functions are used in conjunction with higher order multipliers on the inequality and equality constraints that bound the neural network input-output properties. The volume of the region of attraction computed is increased compared to other methods, allowing for better safety guarantees on the stability of the system. We are also able to easily analyse non-linear polynomial systems, which is not possible to do with other methods. We are also able to conduct robustness analysis on the parameter uncertainty. We show the benefits of our method using numerical examples.
\end{abstract}

\section{Introduction}
Neural networks (NNs) have seen an increase in use in many application areas ranging from weather prediction \cite{czhang} to natural language processing \cite{tbro}. An NN's ability to act as a universal function approximator, combined with the use and availability of big data has meant that they have seen great success in many industries. There is work dating back to the 1990s that aims to use an NN as a \emph{controller} in a feedback control loop \cite{wmiller}, \cite{alev}. We will refer to dynamical systems that contain an NN as a feedback controller as Neural Feedback Loops (NFLs). 

Using NNs as controllers has the potential to provide improved performance over traditional approaches. There is also the possibility that they can be used with data-based control methods, to move away from requiring an accurate model of a system. Work from \cite{ffab} presents a general procedure to design NN controllers that preserve the desirable properties of a model predictive control scheme.

However, one of the primary issues with NNs is that they have been shown to be sensitive to adversarial attacks \cite{shua}, which makes their use in safety-critical applications challenging. This is one of the most significant shortcomings of NNs and if we are to use them as controllers then it is vitally important that these issues are overcome. Determining robustness certificates on NFLs such as stability and reachability is  important if they are to be used more frequently in practice. 

A primary reason why NNs can be hard to verify as safe is due to the non-linearities that are included in the activation functions, which are the elements that have made NNs so successful. One approach that has been used to verify NNs is to bound the non-linearities with appropriate equality and inequality constraints \cite{hsal}. There are many methods that can achieve this, each often trading off solution accuracy for computational time. The choice of bounds (linear, quadratic etc.) influence the type of optimisation problem formulated and hence the computational complexity of the problem. Initial works to quantify the safety of an NFL include \cite{ktan}, which looks at the Lyapunov stability using linear differential inclusions. More recent works have looked at absolute stability by using a transformation into non-linear operator form \cite{kkim}. By using a reinforcement learning perspective, \cite{mjin} were able to get robust stability guarantees. Asymptotic stability is computed along with a region of attraction (ROA) in \cite{hyin} using Lyapunov stability theory, with local sector constraints on the non-linearities. This is taken further by using Integral Quadratic Constraints (IQCs) to bound perturbations in the plant model. It is possible to combine this framework with imitation learning \cite{hyin2}, by training an NN controller with stability guarantees. Another approach that uses IQCs does so in conjunction with the circle criterion and Zames-Falb multipliers \cite{ppau2} to obtain better stability guarantees and larger ROAs. These ideas can be applied to recurrent NNs as shown in \cite{yebi}.

Stability analysis is not the only robustness certificate that can be used to analyse NFLs. Reachability has also been applied which uses similar approaches to bound the activation functions. \cite{mfaz2} uses quadratic constraints with a semi-definite programming framework to obtain tight bounds on the reachable sets. By partitioning the input space efficiently and solving multiple optimisation problems, \cite{mever} computes tighter bounds on the reachable sets in significantly reduced computational time. Bernstein polynomials are used to verify NNs with activation functions of a general form to effectively compute the reachable sets \cite{chuan}.  

We refer to NFLs that have a non-linear plant as non-linear NFLs. We note that previous methods that verify stability consider systems with linear plant models or abstract the non-linearities and bound them using sector or slope constraints. This method has been shown to be effective, however it is possible to obtain better results by considering the non-linear system directly. In this paper, we use sum of squares (SOS) programming to directly incorporate the non-linearities of the plant, instead of relying on such bounds. Another advantage of this method is that we are able to use higher-order multipliers in the Lyapunov constraints, which usually results in better tests. We consider the system in continuous time, as it does not require us to discretise the system and introduce approximations/conservativeness into the framework. 

In Section \ref{sec:nflsummary} we introduce NFLs and how their input-output properties can be represented as a semi-algebraic set. Then in Section \ref{sec:nflstability} we describe how stability can be framed as an optimisation problem and then how it can be verified, along with how an ROA can be computed using SOS programming. We show the results of our method with numerical examples in Section \ref{sec:numexamples}. The paper is then concluded in Section \ref{sec:stabilityconclusion}, outlining plans for future work.

\section{Neural Feedback Loops} \label{sec:nflsummary}
Consider a continuous-time system
\begin{align} \label{eq:nfl2}
\begin{split}
    \dot{z}(t) &= f(z(t),u(t)), 
\end{split}
\end{align}
where $f$ is the plant model (assumed to be a Lipschitz function), $z(t) \in \mathbb{R}^{n_{z}}$, $u(t) \in \mathbb{R}^{n_{u}}$ and $y(t) \in \mathbb{R}^{n_{y}}$ are the system, input and output states respectively and $n_{z}$, $n_{u}$ and $n_{y}$ are the number of system, input and output states respectively. Here $\mathbb{R}^{n}$ denotes the $n$-dimensional real vectors and $\mathbb{R}^{m \times n}$ denotes the set of $m \times n$-dimensional real matrices. A system is considered as a Neural Feedback Loop (NFL) if the controller $u$ is described as a neural network (NN). We consider a state feedback controller $u(t) = \pi(z(t)) \: : \: \mathbb{R}^{n_{z}} \rightarrow \mathbb{R}^{n_{u}}$ as a feed-forward fully-connected NN such that
\begin{align} \label{eq:nnconstab}
\begin{split}
    x^{0}(t) &= z(t), \\ 
    v^{k}(t) &= W^{k}x^{k}(t) + b^{k}, \: \mathrm{for} \: k = 0,\dots, \ell - 1, \\ x^{k+1}(t) &= \phi (v^{k}(t)), \: \mathrm{for} \: k = 0,\dots, \ell - 1, \\
     \pi(z(t)) &= W^{\ell}x^{\ell}(t) + b^{\ell}, 
    \end{split}
\end{align}
where $W^{k} \in \mathbb{R}^{n_{k+1} \times n_{k}}$, $b^{k} \in \mathbb{R}^{n_{k+1}}$ are the weights matrix and biases of the $(k+1)^{th}$ layer respectively and $z(t) = x^{0}(t) \in \mathbb{R}^{n_{z}}$ is the input into the NN. The activation function $\phi$ is applied element-wise to the $v^{k}(t)$ terms. The number of neurons in the $k^{th}$ layer is denoted by $n_{k}$. 

We denote the equilibrium of the states in the closed-loop system as $z^{*}, u^{*}, x^{*}, v^{*}$. For simplicity we will drop the time dependency notation throughout this paper. 

\subsection{Abstracting the Neural Network as Constraints}
A common method for dealing with the non-linearities inside the activation functions is to abstract them as inequality and equality constraints. This method has already been used in many works related to the robustness of NFLs \cite{hyin}, \cite{ppau2}, \cite{mfaz3}. These constraints can be of any form, however common choices are box, slope and sector constraints due to their ability to be expressed in a semi-definite programming framework. In our framework, we are able to choose any set of polynomial constraints to represent the NN, as shown in previous work \cite{mnew2}, \cite{mnew4}. These polynomial constraints can be represented as a semi-algebraic set, which we express with notation
\begin{multline} \label{eq:Sset2}
    S = \big\{ x \in \mathbb{R}^{n} \: | \:  g_{i}(x) \geq 0, \: h_{j}(x) = 0 \: \\ \forall \: i = 1, \dots, p, \: j = 1, \dots , q \big\}, 
\end{multline}
where $g_{i}$ and $h_{j}$ are polynomial functions. Throughout we define $\mathbb{R}[x_{1}, \dots , x_{n}]$ to be the set of polynomials in $x_1, \ldots, x_n$ with real coefficients. We denote $x = (x_1, \ldots, x_n)$ for simplicity.

We now describe briefly how the set $S$ in (\ref{eq:Sset2}) can be obtained. Here, we consider ReLU and tanh activation functions. The ReLU function given by $\mathrm{ReLU}(x) = \phi(x) = \mathrm{max}(0,x)$, can be bound using two inequalities and one equality constraint \cite{arag} such that
\begin{equation*} \label{eq:relu}
    \phi \geq 0, \: \phi - x \geq 0, \: \phi(\phi - x) = 0.
\end{equation*}
We also use a pre-processing step known as interval bound propagation (IBP) \cite{sgow} to compute approximate lower and upper bounds on each activation function $(\underbar{$\phi$}, \overline{\phi})$. These can be expressed as box constraints
\begin{equation*} \label{eq:ibp}
    \phi - \underbar{$\phi$} \geq 0, \: -\phi + \overline{\phi} \geq 0.
\end{equation*}
For the tanh activation function we use sector constraints. It is possible to use the IBP values to create a tight single sector bound as shown in \cite{hyin} and expressed as 
\begin{equation*}
    (\phi - \alpha x)(x - \phi) \geq 0,
\end{equation*}
where $\alpha$ is determined by the IBP bounds. Alternatively we can use two overlapping sectors which can provide better constraints on the activation function when the IBP bounds are large \cite{mnew2}.

It is also possible to use slope constraints to bound the activation functions. Slope constraints are obtained by considering the slope of two activation functions together. This method is employed in \cite{hyin} and \cite{ppau2}, where the slope can be bounded by a sector constraint such that
\begin{equation*}
    \alpha \leq \frac{\phi(x_{2}) - \phi(x_{1})}{x_{2} - x_{1}} \leq \beta.
\end{equation*}
Therefore, any two nodes in the NN must satisfy
\begin{equation*}
    (\phi(x_{i}) - \phi(x_{j}) - \alpha (x_{i} - x_{j}))( \beta (x_{i} - x_{j}) - (\phi(x_{i}) - \phi(x_{j})) \geq 0,  
\end{equation*}
$\forall i,j = 1, \dots, n, \: i \neq j$. For the activation functions in this paper, the slope restricted sectors are $\alpha = 0$ and $\beta = 1$ for ReLU and tanh. However, one issue with slope constraints is that they do not scale well. The number of constraints increases with order $n \choose 2$, as the number of nodes in the NN increases. 

\section{Stability of Neural Feedback Loops} \label{sec:nflstability}
In this paper we aim to show that the equilibrium of the closed loop NFL is stable and then also use this to compute an inner approximation of the region of attraction (ROA). 
\subsection{Stability Conditions}
We first recall the Lyapunov stability theorem \cite{jslot}. We will assume that $z^{*} = 0$ throughout this paper without loss of generality.
\begin{theorem} \label{theorem:Lyapunov}
    Let $z^* = 0$ be an equilibrium point for $\dot z = f(z)$ where $f: \mathbb{R}^n \rightarrow \mathbb{R}^n$ is a Lipschitz function, and $D \subset \mathbb{R}^{n}$ be a domain containing $z=0$. Let $V \: : \: D \longrightarrow \mathbb{R}$ be a continous differentiable function, such that
    \begin{gather*}
        V(0)=0 \: \mathrm{and} \: V(z) > 0 \: \mathrm{in} \: D-\{0\},
        \dot{V}(z) \leq 0 \: \mathrm{in} \: D.
    \end{gather*}
    Then, $0$ is stable. Moreover, if
    \begin{equation*}
        \dot{V}(z) < 0 \: \mathrm{in} \: D-\{0\}, 
    \end{equation*}
    then $0$ is asymptotically stable. 
\end{theorem}
An NFL can be viewed as a dynamical system with equality and inequality constraints that arise from the input-output description of the NN (\ref{eq:Sset2}). In this case, the following theorem from \cite{apap1} can be used to assess the stability of the NFL.
\begin{theorem} \label{theorem:region}
    (\cite{apap1}) Consider the nonlinear system
    \begin{equation*}
        \dot{z} = f(z,u)
    \end{equation*}
    where $z \in \mathbb{R}^{n_{z}}$ is the system state and $u\in \mathbb{R}^{n_{u}}$ denotes a collection of auxiliary variables (parameters, inputs etc). Let the system have the following constraints:
    \begin{align*}
        a_{i_{1}}(z,u) &\geq 0, \: \forall i_{1} = 1, \dots , N_{1}, \\
        b_{i_{2}}(z,u) &= 0, \: \forall i_{2} = 1, \dots , N_{2}, \\
        \int_{0}^{T} c_{i_{3}}(z,u) &\geq 0, \: \forall i_{3} = 1, \dots , N_{3}, \: \mathrm{and} \: \forall T \geq 0,
    \end{align*}
    where $a_{i_{1}}$, $b_{i_{2}}$, $c_{i_{3}}$ are polynomial functions in $(z,u)$ and $f(z,u)$ is a polynomial vector field. The region $D \subset \mathbb{R}^{n_{z}+n_{u}}$ is defined as 
    \begin{equation*}
        D = \{ (z,u)\in \mathbb{R}^{n_{z}+n_{u}} | a_{i_{1}}(z,u) \geq 0, b_{i_{2}}(z,u) = 0, \: \forall i_{1},i_{2}\}.
    \end{equation*}
    Suppose that for the above system there exist polynomial functions $V(z), \: p_{i_{1}}(z,u), \: q_{i_{2}}(z,u)$ and constants $r_{i_{3}}$ such that:
    \begin{itemize}
        \item $V(z)$ is positive definite in a neighborhood of the origin.
        \item $p_{i_{1}}(z,u) \geq 0$ in $D$.
    \end{itemize}
    Then
    \begin{dmath*}
        -\frac{\partial V}{\partial z}f(z,u) - \sum_{i_{1}}p_{i_{1}}(z,u) a_{i_{1}}(z,u) \dots \\
        -\sum_{i_{2}}q_{i_{2}}(z,u) b_{i_{2}}(z,u) - \sum_{i_{3}}r_{i_{3}}c_{i_{3}}(z,u) \geq 0
    \end{dmath*}
    will guarantee that the origin of the state space is a stable equilibrium of the system.
\end{theorem}

We note here that the $u$'s in the theorem above can be the $x$'s describing the NN, extra variables needed to reformulate the system, or parametric uncertainty. Once we have constructed a Lyapunov function showing asymptotic stability of the equilibrium, we can attempt to obtain an estimate of the ROA of that equilibrium.
\subsection{Region of Attraction Conditions}
\begin{definition}
    The region of attraction (ROA) $\mathcal{R}$ of System \eqref{eq:nfl2} is defined as \begin{equation*}
        \mathcal{R} := \{ z(t) \in \mathbb{R}^{n_{z}} \: : \: \lim_{t \to \infty} (z(t)) = 0 \}.
    \end{equation*}
\end{definition}
One can obtain an inner approximation to $\mathcal{R}$, which we will denote by $\hat{\mathcal{R}}$, using the level sets of the Lyapunov function $V(z)$ computed via Theorem \ref{theorem:Lyapunov}. This is achieved by searching for a $\gamma > 0$ such that $\hat{\mathcal{R}} = \{ z\in \mathbb{R}^{n_{z}} | V(z) \leq \gamma \} \subset D$.
\begin{theorem} \label{theorem:ROA}
For $\dot z = f(z)$ suppose that a Lyapunov function $V(z)$ satisfying the conditions of asymptotic stability was constructed, as per Theorem~\ref{theorem:Lyapunov} inside a domain $D$. Suppose $\{z | d(z) \geq 0 \} \subset D$.
If
\begin{equation*}
   |z|^{2k} ( V(z) - \gamma) + p(z)d(z) \geq 0,
\end{equation*}
where $p(z) \geq 0$ and $k$ is an integer, then the level set $V(z)\leq \gamma$ is contained inside the ROA.
\end{theorem}
In the case of a constrained dynamical system, the Lyapunov function constructed via Theorem \ref{theorem:region} can also be used to provide estimates of the ROA. For more details, see~\cite{jand}.

\subsection{Sum of Squares and Stability Analysis}
To construct  the Lyapunov functions presented in the earlier theorems, we use sum of squares (SOS) programming.
\begin{definition} \label{def:sos}
    A polynomial $p(X)$ is a sum of squares (SOS) polynomial if and only if it can be expressed as 
    \begin{equation*}
        p(X) = \sum_{i=1}r_{i}^{2}(X) \equiv p(X) \: \mathrm{is} \: \mathrm{SOS}.
    \end{equation*}
    We define the set of polynomials that admit this decomposition by $\Sigma[X]$.
\end{definition}
Since $p(X) \in \Sigma[X]$ implies that $p(X) \geq 0$, we can replace the non-negativity conditions in Theorem \ref{theorem:region} with SOS conditions and construct these Lyapunov functions using SOS programming, which is equivalent to semi-definite programming. The following subsections explain how this is achieved. 

\subsection{Neural Feedback Loop Stability Conditions}
The global Lyapunov stability SOS conditions for an unconstrained system $\dot{z} = f(z)$ (see Theorem~\ref{theorem:Lyapunov}) are
\begin{align*}
    &V(z) - \rho(z) \in \Sigma[z], \\
    &-\frac{\partial V}{\partial z} f(z) \in \Sigma[z], 
\end{align*}
where $\rho(z)$ is positive definite. 

Similarly, we can write the Lyapunov stability conditions as SOS stability conditions for System \eqref{eq:nfl2}, in feedback with the controller (\ref{eq:nnconstab}).
\begin{proposition} \label{prop:1}
Consider System (\ref{eq:nfl2}) in feedback with a NN controller given by (\ref{eq:nnconstab}). Suppose the input-output properties of the NN are described by (\ref{eq:Sset2}). Suppose there exists a polynomial function $V(z)$ satisfying the following conditions
\begin{align} \label{sosopt1}
\begin{split}
    &V(z) - \rho(z) \in \Sigma[z], \\
    &-\frac{\partial V}{\partial z} f(z,\pi(z)) - \sum_{j}^{q}t_{j}(X)h_{j}(x) \dots \\
    &- \sum_{i}^{p}s_{i}(X)g_{i}(x) \in \Sigma[X], \\
    &\rho(z) >0 , \\
    &s_{i}(X) \in \Sigma[X], \: \forall i = 1, \dots, p, \\ 
    &t_{j}(X) \in \mathbb{R}[X], \: \forall j = 1, \dots, q,
\end{split}
\end{align}
where $X$ is a vector of all the system and NN states, i.e. $X = (x,z)$. Then the equilibrium of the NFL is stable. 
\end{proposition}

The equality and inequality constraints in the SOS program above include the set of constraints from the NN abstraction (\ref{eq:Sset2}) and form a semi-algebraic set. Further constraints may be needed when only local asymptotic stability is being verified. In this case, we describe a region 
\begin{equation}
    D^{z} = \left \{ z \in \mathbb{R}^{n_z} | d_{k}(z) \geq 0, \quad k=1,\dots, n_{d} \right \}. \label{eq:Dz}
\end{equation}
where the stability conditions will need to be satisfied. These constraints can be incorporated into the optimisation framework in the same way as in Theorem \ref{theorem:region}.
\begin{proposition}
Consider System (\ref{eq:nfl2}) in feedback with a NN controller given by (\ref{eq:nnconstab}). Suppose the input-output properties of the NN are described by (\ref{eq:Sset2}), and consider the region given by (\ref{eq:Dz}). Suppose there exists a polynomial function $V(z)$ satisfying the following conditions
\begin{align} \label{sosopt2}
\begin{split}
    &V(z) - \rho(z) \in \Sigma[z], \\
    &-\frac{\partial V}{\partial z}(z) f(z,\pi(z)) - \sum_{k=1}^{n_d}p_{k}(X)d_{k}(z)  \dots \\
    &- \sum_{j}^{q}t_{j}(X)h_{j}(x)  - \sum_{i}^{p}s_{i}(X)g_{i}(x) \in \Sigma[X], \\
    &\rho(z) >0, \\
    &p_{k}(X) \in \Sigma[X], \: \forall k = 1, \dots , n_{d}, \\
    &s_{i}(X) \in \Sigma[X], \: \forall i = 1, \dots, p, \\
    &t_{j}(X) \in \mathbb{R}[X], \: \forall j = 1, \dots, q. 
\end{split}
\end{align}
Then the equilibrium of the NFL is stable.
\end{proposition}
The above optimisation problems have the benefit of analysing the system in the continuous time domain, whereas methods such as \cite{mkor} only analyse the discrete time system. By using the continuous time domain we are able to ensure stability without discretising the system, which may introduce conservativeness as it acts as an approximation to the true continuous-time system.

Suppose now that a Lyapunov function is constructed using \eqref{sosopt2}. To determine the ROA as stated in Theorem \ref{theorem:ROA} we must determine the largest level set of the Lyapunov function $V(z)$ that is contained within the region that the Lyapunov conditions are satisfied, which can also be cast as an SOS program. Suppose $d(z) \geq 0$ is included inside this region. Then  
\begin{align} \label{eq:sosroa}
\begin{split}
    &|z|^k(V(z) - \gamma) + p(z)d(z)  \in \Sigma[z], \\
    &p(z) \in \Sigma[z],
\end{split}
\end{align}
where $\gamma$ is a variable to be maximised and $k$ is a positive integer, ensures that $V(z) \leq \gamma$ is an estimate of the ROA. This process can then be repeated by adjusting the $d_{k}(z)$ constraints to expand the region tested for stability - we provide an algorithm in the next section. 

\subsection{Algorithm for Computing the Stability and Region of Attraction of a Neural Feedback Loop} 
To compute the stability of System \eqref{eq:nfl2} we must first ensure that the vector field is polynomial. If the system is not already expressed as a polynomial, this can be achieved by making an appropriate substitution \cite{apap1}. We assume that we are given an NN controller for the system and want to establish the stability properties of the closed loop system and obtain an estimate of the ROA. The SOS constraints that are constructed can be solved with a appropriate parser such as SOSTOOLS \cite{sostools} in MATLAB or SumOfSquares.jl in Julia \cite{sosjl}. The procedure for determining stability is as follows:
\begin{algorithm}
\caption{Determining the Stability of an NFL} \label{alg:nflstability}
\begin{algorithmic}
\State Input: NFL \eqref{eq:nfl2}, NN \eqref{eq:nnconstab}, region (\ref{eq:Dz})
\State Initialise $V(z), p_{k}(X), h_{j}(X), s_{i}(X), \rho(z)$, as in \eqref{sosopt2}
\State $\mathrm{feas} = 0$
\While{$\mathrm{feas} = 0$}
\State Compute IBP values using region (\ref{eq:Dz})
\State Obtain $g_{i}(x)$ and $h_{j}(x)$ from \eqref{eq:nnconstab} and IBP values 
\State Solve \eqref{sosopt2}
\If{\eqref{sosopt2} is feasible}
    \State $\mathrm{feas} \gets  1$
\Else{}
    \State $d_{k} \gets \hat{d}_{k}$ \Comment{$\hat{d}_{k}$ defines a new region $\hat{D}$}
\EndIf
\EndWhile
\State Output: $V(z)$
\end{algorithmic}
\end{algorithm}

We now describe these steps in detail. Before the NN constraints can be created, we must first define the region $D^z$ of the state space that we are interested in by setting the constraints $d_{k}$. Once these are established, we can then compute the maximum and minimum possible values of each system state $(\underbar{$z$}, \overline{z})$. These values can be used as inputs into the IBP algorithm to find the preprocessing bounds. Once these bounds have been computed the NN constraints can be obtained to form the semi-algebraic set \eqref{eq:Sset2}. These constraints will depend on the problem set-up. Having established the constraints that define our system, we search for a Lyapunov function using \eqref{sosopt2}. If this is infeasible, we can reduce the size of the region $D^z$ until a Lyapunov function can be found. Once a Lypaunov function is found the ROA can be obtained using \eqref{eq:sosroa}.

\subsection{Extension to Robustness Analysis}
So far, we have assumed that the parameters within our system are fixed. When there is parametric uncertainty, we wish to ensure robust stability. Theorem~\ref{theorem:region} allows for parameters to be included in the description of the system and therefore it is possible to construct parameterised Lyapunov functions. We demonstrate this in an example below. 

\section{Numerical Examples} \label{sec:numexamples}
We demonstrate our approach on various systems to show improvements over existing methods. All experiments were run on a 4-core Intel Xeon processor @3.50GHz with 16GB of RAM. We refer to our method as `NNSOSStability'. We implement our method using MATLAB and SOSTOOLS to parse the SOS constraints into a semi-definite program, which is solved using SeDuMi \cite{sedumi}.

\subsection{Duffing Oscillator} \label{sec:duffosc}
\begin{figure}[h] 
    \centering  
    \includegraphics[height=6cm]{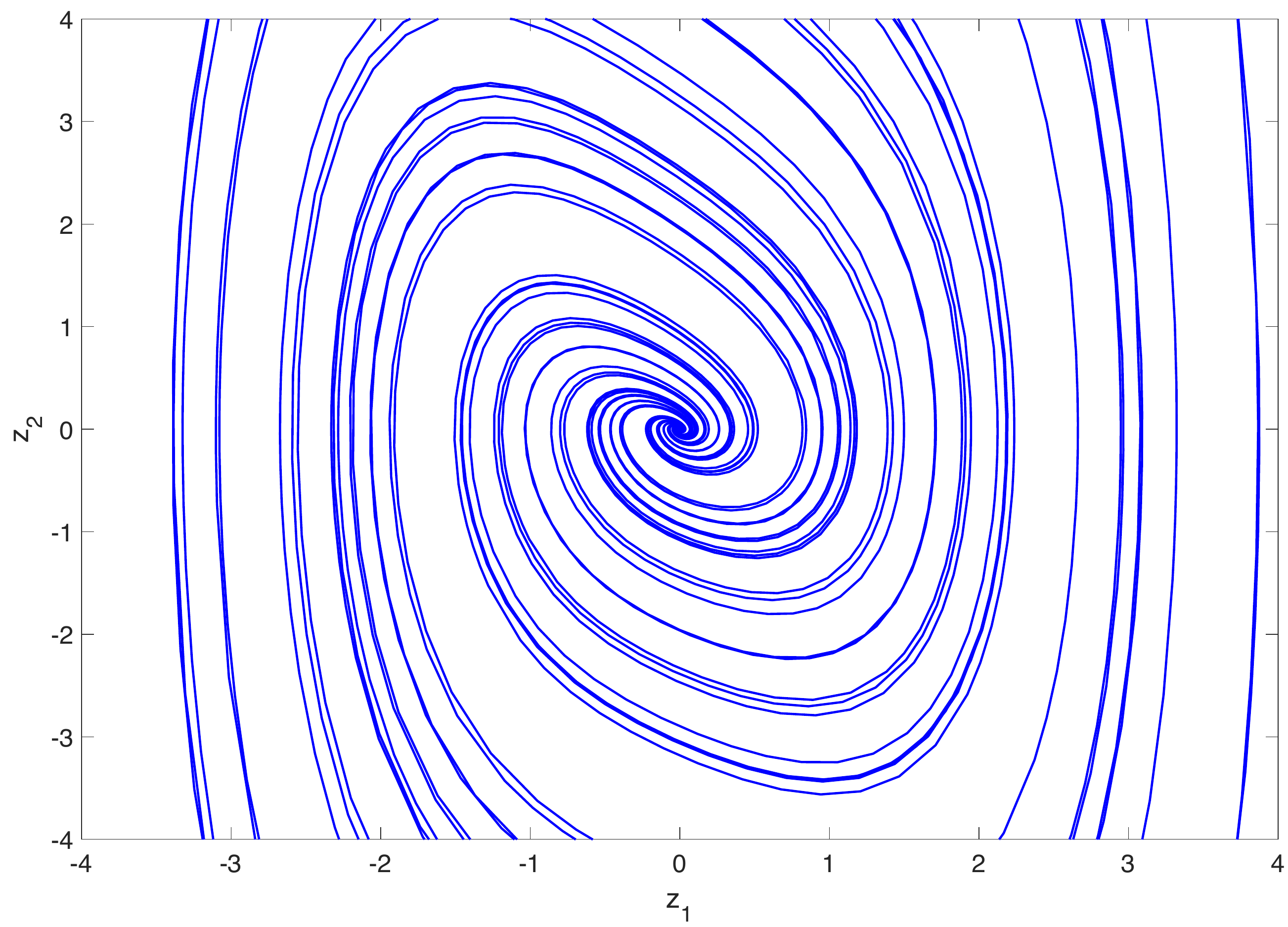}
    \caption{The trajectories of the Duffing Oscillator Example in Section \ref{sec:duffosc}. }  \label{fig:duffosc}
\end{figure}

We show the benefits of this SOS method on a simple non-linear system example. Consider the Duffing Oscillator proposed in \cite{mever}, with
system dynamics
\begin{align*}
    \dot{z}_{1} &= z_{2}, \\
    \dot{z}_{2} &= -z_{1} - 2\zeta z_{2} - z_{1}^{3} + u,
\end{align*}
where $\zeta = 0.5$ is the damping ratio. We train a small NN controller with two layers and two nodes in each layer with ReLU activation functions. We are able to verify global asymptotic stability of this system using a fourth-order Lyapunov function and second-order multipliers. The benefits of this SOS framework is that we are able to account for the non-linearity directly, instead of bounding it by a sector constraint or by making a substitution to lift the variables. For comparison, if we use the linearised model with sector constraints on $-z_{1}^{3}$ we are only able to verify stability in a circle of radius 0.2. If we use the discrete Lyapunov stability conditions then the SOS program becomes too large and we cannot verify global stability. The trajectories of this system are shown in Figure \ref{fig:duffosc}.

\subsection{Three Dimensional Polynomial System} \label{sec:example3d}
\begin{figure}[h] 
    \centering  
    \includegraphics[height=6cm]{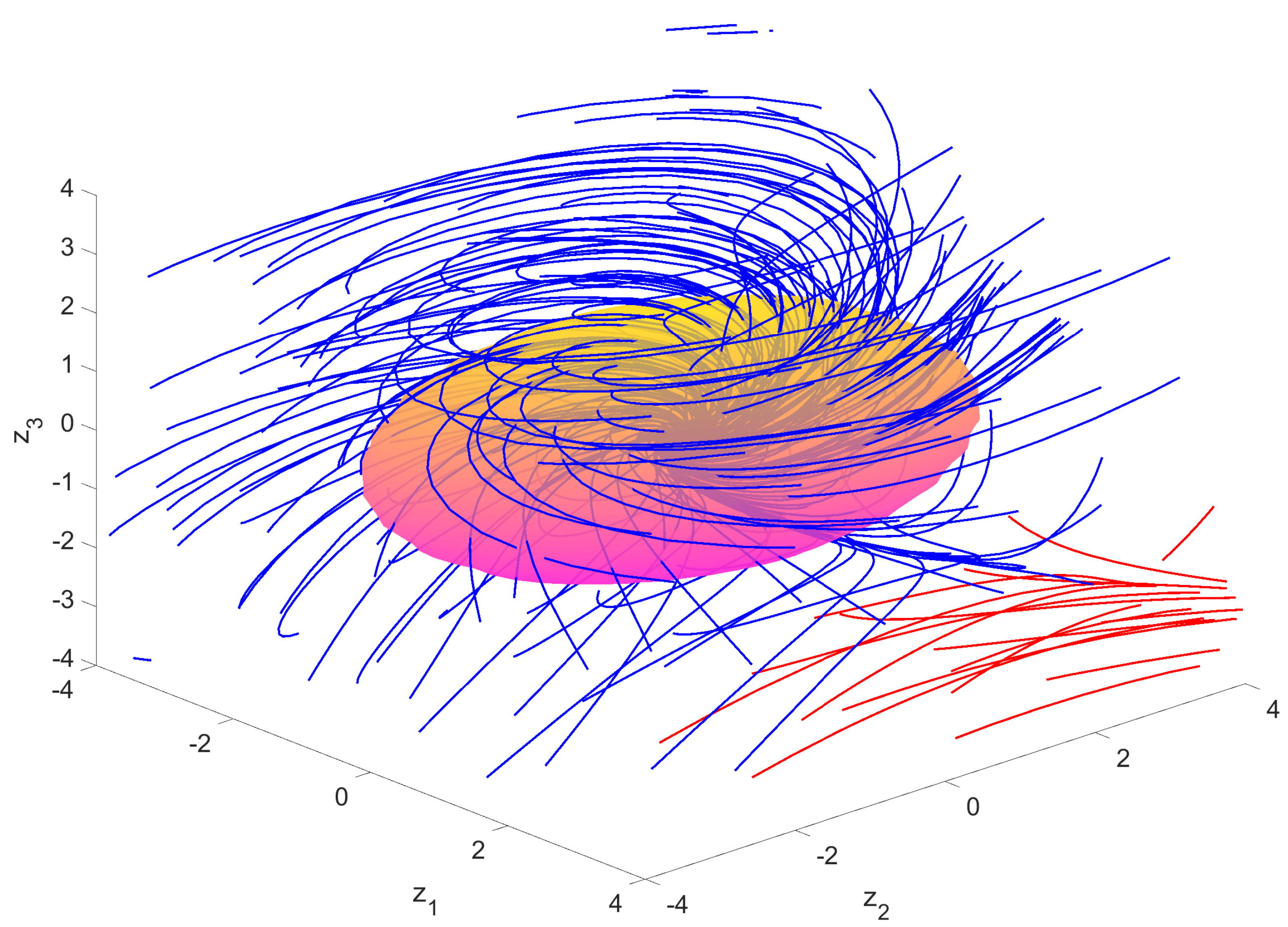}
    \caption{ROA of NNSOSStability in Example \ref{sec:example3d} with a second-order Lyapunov function shown as a surface. The blue trajectories tend towards the zero equilibrium, whereas the red ones do not.}  \label{fig:newsys3d}
\end{figure}

\begin{figure}[h] 
    \centering 
    \includegraphics[height=6cm]{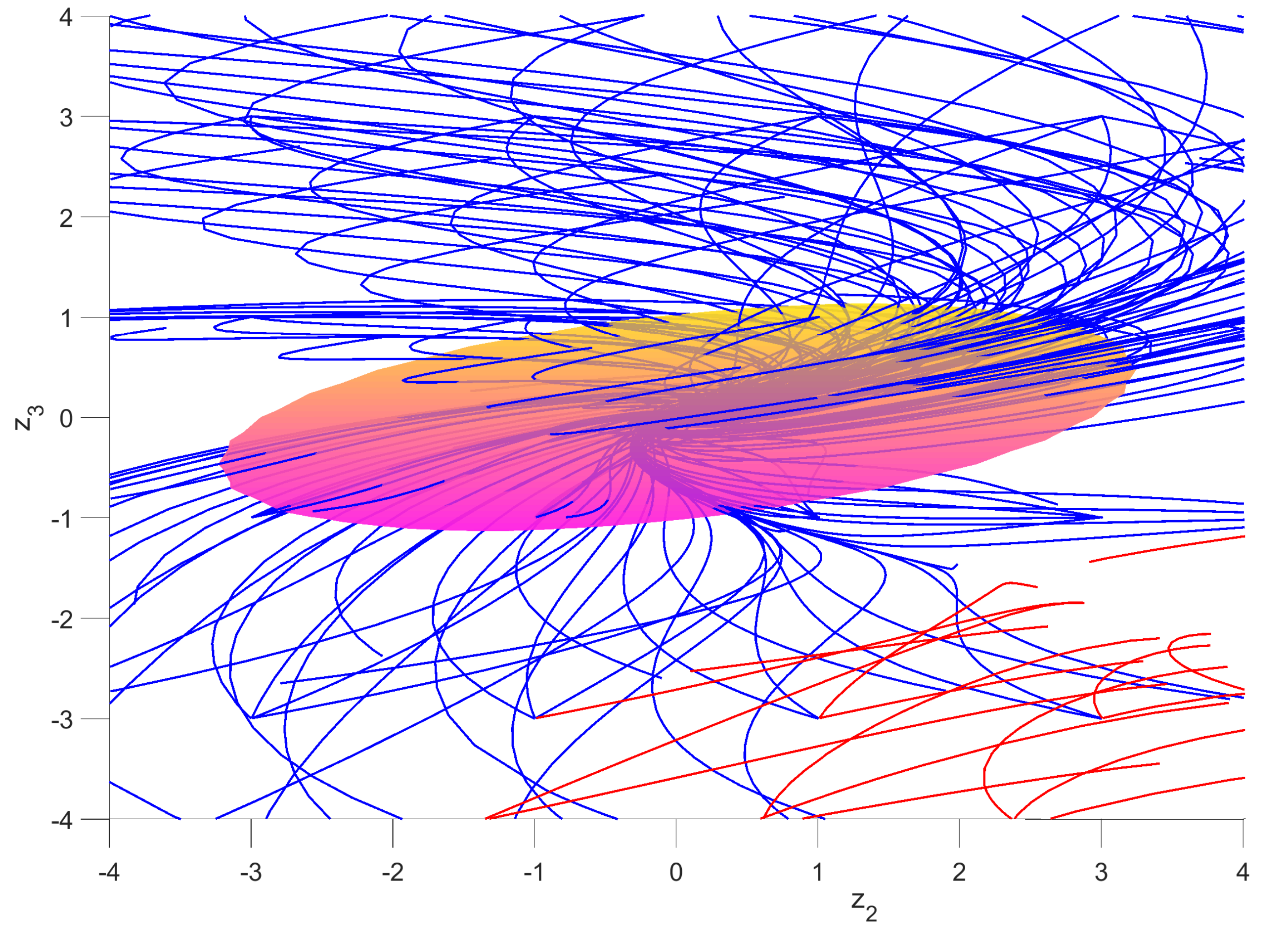}
    \caption{ROA of NNSOSStability in Example \ref{sec:example3d} with a second-order Lyapunov function shown as a surface. The blue trajectories tend towards the zero equilibrium, whereas the red ones do not.}  \label{fig:newsys2d}
\end{figure}

The previous example shows that the stability of NFL systems can be verified effectively using SOS and demonstrates the benefits it possesses over other methods. However, the uncontrolled system is globally stable. We now consider a non-linear polynomial plant with three states that is unstable without a controller. The system is given by
\begin{align*}
    \dot{z}_{1} &= -z_{1} + z_{2} - z_{3}, \\
    \dot{z}_{2} &= -z_{1}(z_{3} + 1) - z_{2}, \\
    \dot{z}_{3} &= -z_{1} + u.
\end{align*}
We train a five layer NN with five nodes in each layer with tanh activation functions to stabilise the system. We specify the region $D$ to be a cube of length six such that
\begin{gather*}
    3 - z_{1} \geq 0, \: 3 - z_{2} \geq 0, \: 3 - z_{3} \geq 0, \\
    3 + z_{1} \geq 0, \: 3 + z_{2} \geq 0, \: 3 + z_{3} \geq 0,
\end{gather*}
to compute the IBP bounds, which are used to compute the relevant sector and slope constraints on the activation functions. We propose a quadratic Lyapunov function, which we use to verify stability. The ROA is found to be a good approximation of the true ROA, which is computed using an exhaustive search. The ROA in three dimensions is shown in Figure \ref{fig:newsys3d} and a rotated view is shown in Figure \ref{fig:newsys2d}.

\subsection{Stability and Region of Attraction Analysis of Inverted Pendulum}
We consider the inverted pendulum example proposed in \cite{hyin}, trained with the NN controller in \cite{ppau2}. The system dynamics are
\begin{gather*}
    \ddot{\theta}(t) = \frac{mgl\sin{(\theta(t))} - \mu \dot{\theta}(t) + \mathrm{sat}(u(t))}{ml^2},
\end{gather*}
where $m=0.15 \: \mathrm{kg}, l=0.5 \: \mathrm{m}, \mu=0.5 \: \mathrm{Nmsrad}^{-1}, g=9.81 \: \mathrm{ms}^{-2}, u_{max} = 1 \: \mathrm{Nm}$ and $\mathrm{sat}(u(t))$ is the saturation function on the control input. To incorporate this system in the SOS programming framework, we let $z_{1} = \theta$, $z_{2} = \dot{\theta}$ and $z_{3} = z_{1} - \mathrm{sin}(z_{1})$. This replaces the non-linear system with a linear system, however we must bound the $z_{3}$ term, which we can do with a sector constraint depending on how our region $D$ is defined. This system is an approximation of the original non-linear system and can be written as 
\begin{align*}
    &\dot{z_{1}} = z_{2}, \\
    &\dot{z_{2}} = \frac{g}{l}(z_{1} - z_{3}) - \frac{\mu}{ml^2} z_{2} + \frac{1}{ml^2} u, \\
    &-u_{max} \leq u \leq u_{max}, \\
    &z_{3}(\alpha_{z_{3}}z_{1} - z_{3}) \geq 0,
\end{align*}
where 
\begin{equation*}
    \alpha_{z_{3}} = \max \bigg( \frac{\overline{z}_{1}  - \mathrm{sin}(\overline{z}_{1})}{\overline{z}_{1}}, \frac{\underbar{$z$}_{1}  - \mathrm{sin}(\underbar{$z$}_{1})}{\underbar{$z$}_{1}} \bigg),
\end{equation*}
is determined by the maximum and minimum bounds of the system states. The input saturation constraints and the sector constraints on the non-linearity are added to the derivative condition in the optimisation problem. The NN has five layers and five nodes in each layer with tanh activation functions. 

\begin{figure}[h] 
    \centering  
    \includegraphics[height=6cm]{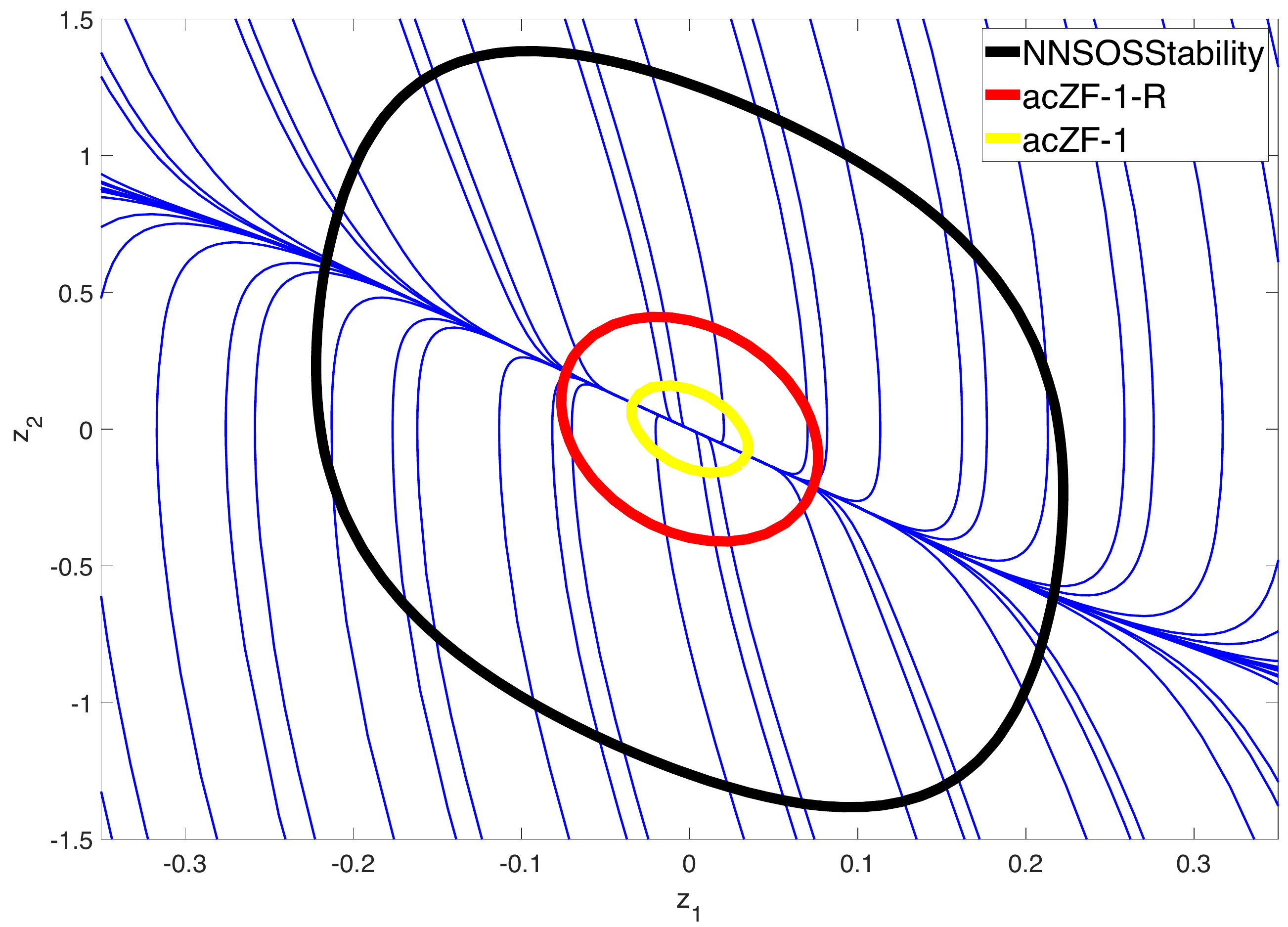}
    \caption{ROA of NNSOSStability with a fourth-order Lyapunov function (black). The ROA for the method in \cite{ppau2} which uses diagonal (yellow) and full-block (red) acausal Zames-Falb multipliers respectively. The trajectories are shown in blue.} \label{fig:IP1}
\end{figure}

We search for a quartic Lyapunov function $V(z_{1},z_{2})$ and compare its performance to other methods. We set the system constraints to be a box such that 
\begin{gather*}
    z_{1} - \underbar{$z$}_{1} \geq 0, \: z_{2} - \underbar{$z$}_{2} \geq 0, \\
    \overline{z}_{1} - z_{1} \geq 0, \:\overline{z}_{2} - z_{2} \geq 0. 
\end{gather*}
We use sector and slope constraints on the tanh activation functions in the NN. Choosing values of $\underbar{$z$}_{1} = -0.3$, $\overline{z}_{1} = 0.3$, $\underbar{$z$}_{2} = -1.4$ and $\overline{z}_{2} = 1.4$ we are able to obtain a fourth-order Lyapunov function in this region. We then compute the ROA for this Lyapunov function and find that its volume is increased significantly over other methods, as shown in Figure \ref{fig:IP1}. We compare this method to the linearised system with different kinds of Zames-Falb multipliers from \cite{ppau2}. We also tried methods from \cite{hyin} and \cite{mkor}, however we were unable to find any solution to the optimisation problem.

\subsection{Robustness Analysis of Inverted Pendulum}
We use the above example and conduct robustness analysis on the system to determine how the stability is affected by perturbations on the length of the pendulum arm. We make the substitution $\delta = 1/l$ and define a robustness certificate on $\delta$. We can rewrite the system as
\begin{align*}
    &\dot{z_{1}} = z_{2}, \\
    &\dot{z_{2}} = g \delta (z_{1} - z_{3}) - \frac{\mu \delta^2}{m} z_{2} + \frac{\delta^2}{m} u, \\
    &-u_{max} \leq u \leq u_{max}, \\
    &z_{3}(\alpha_{z_{3}}z_{1} - z_{3}) \geq 0, \\
     &\Delta(\delta) \geq 0,
\end{align*}
where $\Delta(\delta)$ is a function describing the robustness of the variable $\delta$. We perturb the length by $\pm 0.3$, which generates the robustness constraint
\begin{equation*}
    (5 - \delta)(\delta - 1.25) \geq 0.
\end{equation*}

We propose a quadratic Lyapunov function with a region defined with values of $\underbar{$z$}_{1} = -0.1$, $\overline{z}_{1} = 0.1$, $\underbar{$z$}_{2} = -0.3$ and $\overline{z}_{2} = 0.3$. We are able to find a feasible solution to the SOS optimisation problem, which shows that the system is stable in that region under the perturbation $\Delta(\delta)$.

\section{Conclusion} \label{sec:stabilityconclusion}
In this paper, we have proposed an SOS programming framework to determine the stability of NFLs in the continuous time domain. We use a well established approach to abstract the NN as a set of equality and inequality constraints to construct a semi-algebraic set. We then use Lyapunov stability theory for constrained systems and SOS programming to construct the relevant Lyapunov certificates. Once a Lyapunov function is constructed within a specified region, an ROA of the equilibrium can be computed by solving another SOS program. This framework allows us to conduct robustness analysis if parametric or other uncertainties are present. We test our method on various numerical examples and compare our results to existing methods. 

There are many future directions that can be taken to improve the results in this paper. The first is by using sparsity exploiting methods proposed in \cite{mnew4} to reduce the computational time required to solve these SOS optimisation problems. This would also allow us to extend these results to larger scale systems. It is also possible to combine this method with other recent approaches to improve performance. For example, finding a way to incorporate state-of-the-art NN verification solvers such as \cite{betacrown}, might be beneficial. Finally we would like to extend this method to focus on the control synthesis problem, by training the NN whilst providing robustness guarantees in the process. 

\printbibliography

\end{document}